\documentclass[jpc,a4paper]{revtex4}

\usepackage[utf8]{inputenc}
\usepackage[frenchb]{babel}
\usepackage{amsmath,soul}
\usepackage{amsfonts}
\usepackage{amssymb}
\usepackage[pdftex]{color,graphicx}

\begin{document}

\author{J. Pedregosa-Gutierrez}\email{jofre.pedregosa@univ-amu.fr}
\affiliation{Aix Marseille Univ, CNRS, PIIM, Marseille, France}
\author{C. Champenois}
\affiliation{Aix Marseille Univ, CNRS, PIIM, Marseille, France}
\author{M. Houssin}
\affiliation{Aix Marseille Univ, CNRS, PIIM, Marseille, France}
\author{M.R. Kamsap}
\altaffiliation{Dept de Physique, Universit\'e des Sciences et Techniques de Masuku, Franceville, Gabon}
\author{M. Knoop}
\affiliation{Aix Marseille Univ, CNRS, PIIM, Marseille, France}

\title{Correcting symmetry imperfections \\in  linear multipole traps}
\keywords{keywords: radio-frequency multipole traps; laser cooling; symmetry breaking; 2D structures}

\date{\today}

\begin{abstract}
Multipole radio-frequency traps are central to collisional experiments in cryogenic environments. They also offer possibilities to generate new type of ion crystals topologies and in particular the potential to create infinite 1D/2D structures: ion rings and ion tubes. However, multipole traps have also been shown to be very sensitive to geometrical misalignment of the trap rods, leading to additional local trapping minima. The present work proposes a method to correct non-ideal potentials, by modifying the applied radio-frequency amplitudes for each trap rod. This approach is discussed for the octupole trap, leading to the restitution of the ideal Mexican-Hat-like pseudo-potential, expected in multipole traps. The goodness of the compensation method is quantified in terms of the choice of the diagnosis area, the residual trapping potential variations, the required adaptation of the applied radio-frequency voltage amplitudes, and the impact on the trapped ion structures. Experimental implementation for macroscopic multipole traps is also discussed, in order to propose a diagnostic method with respect to the resolution and stability of the trap drive. Using the proposed compensation technique, we discuss the feasibility of generating a homogeneous ion ring crystal, which is a measure of quality for the obtained potential well.
\end{abstract}


\maketitle
\section{Introduction}
The radio-frequency (RF) ion trap is largely used in physics and chemistry: from quantum computing~\cite{friis18}, metrology~\cite{chen17}, and the observation of time-crystals \cite{zhang17}, to the study of large ion crystals \cite{hornekaer02,kamsap15a}. While the quadrupole configuration is the most widely used type of linear trap, multipole traps are a common choice for cold collision experiments due to their lower RF-heating~\cite{gerlich92}. Numerical simulations of laser-cooled ions in multipole RF  traps have been shown to lead to Coulomb crystals with new topologies: ion rings and ion tubes~\cite{okada07,champenois10,marciante12}. These original ion arrangements with respect to the quadrupole trap are the consequence of an electric field in the trap centre with higher-order terms.

With their higher degree of symmetry, ion rings could outperform ion chains for a number of applications (see examples in~\cite{pedregosa18}), in particular they have been proposed for applications of high-resolution spectroscopy~\cite{champenois10}.

Ion rings have been created experimentally in a quadrupole storage ring~\cite{waki92}. Micro-fabricated surface traps with a large number of electrodes have allowed to trap up to 400 ions in a ring structure with a radius of 624~$\mu$m~\cite{tabakov15} (with an average ion separation of 9~$\mu$m), or up to 15~ions in a 90~$\mu$m-radius ring~\cite{li17}.

Multipole RF traps have been used for several decades in order to trap cold ion clouds~\cite{gerlich92}. The existence of local minima in multipole traps was first reported by Otto et al. \cite{otto09} in the trapping potential of a 22-pole trap, as evidenced by the spatial distribution of H$^{-}$ ions at a temperature of 170~K. The same group has also reported incomplete cooling of trapped ions \cite{endres17}, which could possibly be a consequence of a non-ideal trapping potential. Numerical simulations in the case of a non-ideal 3D octupole and a 22-pole trap have reported potential minima with a local depth in the $meV$-range \cite{fanghanel2017}.

Very recently, we have reported experimental evidence of three local minima in an octupole trap~\cite{pedregosa18}. Extensive numerical characterisation of these minima was also performed. It was found that extremely small deviations from the ideal trap geometry - as small as a rod position error of $0.2\%$ relatively to the trap radius - are enough to lead to observable $(k-1)$ local minima, where $2k$ is the number of trap rods~\cite{pedregosa18}. These results demonstrate the need for a compensation mechanism in order to eliminate potential inhomogeneities and ultimately lead to the experimental creation of ion rings in linear multipole traps. In this paper we propose a compensation strategy that allows to correct the trapping potential in a macroscopic multipole RF trap and we discuss its goodness in terms of various parameters. As it can be understood as a sequence of 2D errors, any lack of parallelism between the rods is not taken into account and we only consider a mismatch of rod positions in a transverse plane.


This article is organised as follows. First, the analytic description of the potential in the multipole trap is introduced. We then present the chosen strategy to implement compensation of potential mechanical defects followed by a numerical study of the performances and limits of this technique. We also perform numerical experiments using molecular dynamics simulations to show that the trapping potential can be corrected to a degree that makes the generation of ion rings in a multipole RF trap indeed experimentally feasible.

\section{Compensation strategy for multipole potentials}
\subsection{Trapping potentials in a linear multipole rf trap}
The radial potential generated by an ideal and infinitely long linear RF trap formed by $2k$ rods alternately polarised by an oscillating electric potential with amplitude $V_{RF}$  and frequency $\Omega/2\pi$  can be written in polar coordinates as~\cite{gerlich92}:
\begin{equation}\label{eq:ideal_octo}
 \phi_{rf}(r,\theta,t) = V_{RF}\cos(\Omega t)\left(\frac{r}{r_{0}}\right)^{k} \cos(k\theta)
\end{equation}
where  $r_{0}$ is the inner radius of the trap. In the adiabatic approximation, the  motion of the ions can be understood as if they were trapped in a static potential known as the pseudo-potential, given by:
\begin{align}\label{eq:pseudo}
\phi_{ps}(r,\theta) = \frac{ Q }{4 m \Omega^{2} } \times \left|\vec{E}(r,\theta)\right|^{2}
\end{align}
with $Q$ and $m$ being the electric charge and the mass of the trapped ions and $\vec{E}$ is the amplitude of the electric field generated by the potential $\phi_{rf}$. Confinement along the trap axis ($z$-axis) is obtained by DC potentials applied to extra electrodes at the end of the trap or outer segments of the rods. It can be approximated by its lowest order expansion
\begin{equation}
\phi_{dc}(r,z) = \frac{1}{2}m\omega_{z}^{2} \left(z^{2} - \frac{r^{2}}{2}\right).
\label{eq:dc_part}
\end{equation}
where $\omega_{z}$ is the oscillator frequency along the $z$-axis with $\omega_{z}^{2}$ proportional to the applied DC potential.

For a multipole trap with $k>2$, the sum of these two potentials forms a trapping potential with the radial profile of a Mexican Hat (MH), due to the non-confining radial contribution of $\phi_{dc}(r,z)$, making the centre of the trap an unstable equilibrium position. Ions settle at the minima of the MH potential, $r_{min}$, only if the Coulomb repulsion potential can be neglected compared to the DC potential \cite{champenois10,marciante12,cartarius13}. When cold enough, ions are expected to form a hollow core structure which can be described as a ring \cite{champenois10,cartarius13}, a mono-layer tube \cite{marciante12}, or a multi-layer tube \cite{calvo09}. This organisation can be understood in the mean-field limit as the condition for a Poisson-Boltzmann equilibrium at low temperature \cite{champenois09}.

\subsection{Implementation of correction}
As introduced before, imperfections of the trap geometry have already lead to the observation of 10 minima in a 22-pole trap~\cite{otto09} and 3 minima in an octopole~\cite{pedregosa18}. In order to experimentally achieve the predicted ring and tube Coulomb crystals, 
any experimental attempt will have to incorporate a compensating mechanism to face the unavoidable mechanical imperfections.

In an experimental context it is suited to express the trap potential as the sum of the potentials generated by each of the rods that form the trap:
\begin{equation}\label{eq:experimental_octo}
\phi_{rf}(x,y,t) = \sum_{l=0}^{2k-1}{ \left[(-1)^{l} V^{RF}_{l}\cos{\Omega t} + V^{DC}_{l}\right] \Phi_l(x,y)}
\end{equation}
where $V^{RF}_{l}$ and $V^{DC}_{l}$ are respectively the RF and DC voltages applied on rod $\l$ and where $\Phi_{l}(x,y)$ corresponds to the potential generated by one rod with 1~V  applied  while all  other electrodes are at zero. The $\Phi_{l}(x,y)$ depend on the particular geometry of the system and it is assumed that they do not change over time.
The potential $V_{sym}$ generated in the symmetrical error-free case corresponds to $V^{RF}_{l}=V_{RF}$ for each rod $\l$, $V^{DC}_{l}=0$ and $\Phi_{l}(x,y)=\Phi^{sym}_{l}(x,y)$ for a  geometry of the rods with perfect  cylindrical symmetry.

To compensate deviations from the error-free geometry, it is clear from the precedent equation that there are only two free parameters per rod: $V^{RF}_{l}$ and $V^{DC}_{l}$. As the local minima appear only in the adiabatic approximation, it is the oscillating part, $V^{RF}_{l}$, that has to be taken as a compensation tool.
Additional static components are often used to compensate possible extra static fields and to reduce excess micro-motion~\cite{berkeland98}. They are not the focus of this work and therefore they are not considered, allowing us to  set $V^{DC}_{l} = 0$. In practice, this only leaves the possibility of using different RF voltage amplitudes for each trap rod. While this is technically challenging, it has already been implemented in a different context~\cite{herskind09}.

Let us now consider how to determine the values of $V^{RF}_{l}$ that best compensate a given misalignment of the trap rods and how to evaluate the success of such a compensation. The approach we use is based on a standard least-squares minimisation technique applied to the difference between the error-free potential
$V_{sym}$, and a potential calculated for a given misalignment of the rods with $V^{RF}_{l}$ as parameters. This can be mathematically expressed as:
\begin{align}\label{eq:chi2}
\chi^{2} = \sum_{i}\sum_{j} \left(V_{sym}(x_{i}, y_{j}) - \sum_{l}{(-1)^{l} V^{RF}_{l} \Phi_{l}(x_{i},y_{j}) } \right)^{2}
\end{align}
where $\chi^{2}$ is the quantity to be minimised.

Due to the linear dependence of the coefficients with respect to $\Phi_{l}(x,y)$, linear algebra can be used to transform the minimisation problem to a matrix inversion problem~\cite{bevington02}.
The choice of the region used to calculate the double sum $\sum_{i}\sum_{j}$ is important and will be discussed later.

The fields have been computed using a commercial Boundary Element Method software\cite{CPO}. The grid resolution $dx$ was modified to verify convergence on the determination of the $V^{RF}_{l}$ values. The following results were obtained with grid resolution of $dx = r_0/2000$. The same resolution was used in the $x$ and $y$ direction.


\begin{figure}
\center
\scalebox{0.65}{\includegraphics{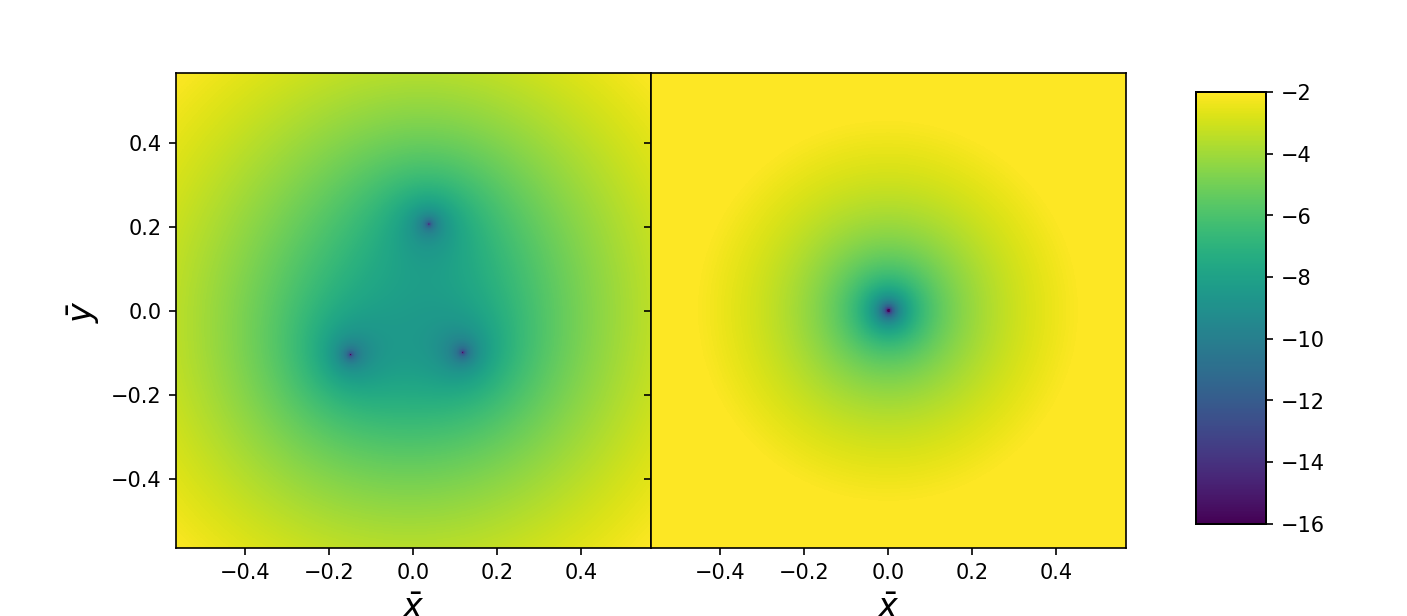}}
\caption{Pseudo-potential (in logarithmic arbitrary units) corresponding to our experimental set-up before (left) and after (right) compensation. Reduced coordinates have been used for the spatial axis ($\bar{x}=x/r_{0}$, $\bar{y}=y/r_{0}$). The same color-scale has been used on both images.}
\label{fig:pseudo_example}
\end{figure}

Once the values of $V^{RF}_{l}$ obtained from the fitting procedure, the resulting compensated pseudo-potential can be computed. Figure~\ref{fig:pseudo_example} shows an example corresponding to the  case of our experimental set-up where Ca$^{+}$-ions are laser-cooled in a linear octupole trap. Trap parameters are: $r_{0}=3.93$ mm, $\Omega/2\pi = 3.5$~MHz, $V_{RF} = 300$~V. The defects have been chosen to reproduce our observations corresponding to a mean random displacement for the rods of $0.013 r_{0} = 51 \mu$m, see~\cite{pedregosa18} for  details. On the left-hand side, Figure~\ref{fig:pseudo_example} shows the initial uncompensated pseudo-potential and on the right the compensated pseudo-potential.
Unless otherwise indicated, we use the indicated trap parameters for all future simulations.

In the next section we discuss the goodness of these compensation results.

\section{Evaluation of the quality of the compensation}

Without compensation, the depth of  these additional potentials is $\approx 50$~K~\cite{pedregosa18} with an average distance to the trap centre of about $<d> = 651 \mu$m.  A closer look at the compensated pseudo-potential in Figure~\ref{fig:pseudo_example} shows that three minima are still present with a distance to the trap centre of $<d> = 4.3\mu$m, and a potential depth evaluated to about $2 \cdot 10^{-12}$~K, which is completely negligible to any other energy scale in the system (for the indicated typical trap parameters). De facto, it can be considered that the compensated potential presents a single minimum.

It is important to remind that among the main interests of such an experimental compensation process is the prospect to achieve a uniformly distributed ion ring structure, or a potential well avoiding additional and uncontrolled ion heating. Therefore the quality of the compensation should be characterised by the degree of homogeneity of the trapping potential along the MH minimum. To evaluate this homogeneity, the contribution of the axial confining electrodes, $\phi_{dc}(r,z)$ has been added to the compensated pseudo potential $\phi_{ps}^{c}$ in order to find the position of the MH  minima and evaluate the total potential at those positions. It is assumed that there is no misalignment of the axial electrodes.

To first-order approximation, the radius of the MH minimum potential, $r_{min}$, in the error-free case can be imposed using equation~(\ref{eq:rmin})~\cite{champenois10}.
\begin{align}\label{eq:rmin}
r_{min}^{4} = \frac{1}{3}\left(\frac{m\Omega \omega_{z}r_{0}^{4}}{2Q V_{RF}}\right)^{2}
\end{align}
In the following, by adjusting $\omega_{z}$ we choose a $\hat{r}_{min} = r_{min}/r_{0}$.

\subsection{In terms of the potential uniformity along the MH minimum}

\begin{figure}
\center
\scalebox{0.55}{\includegraphics{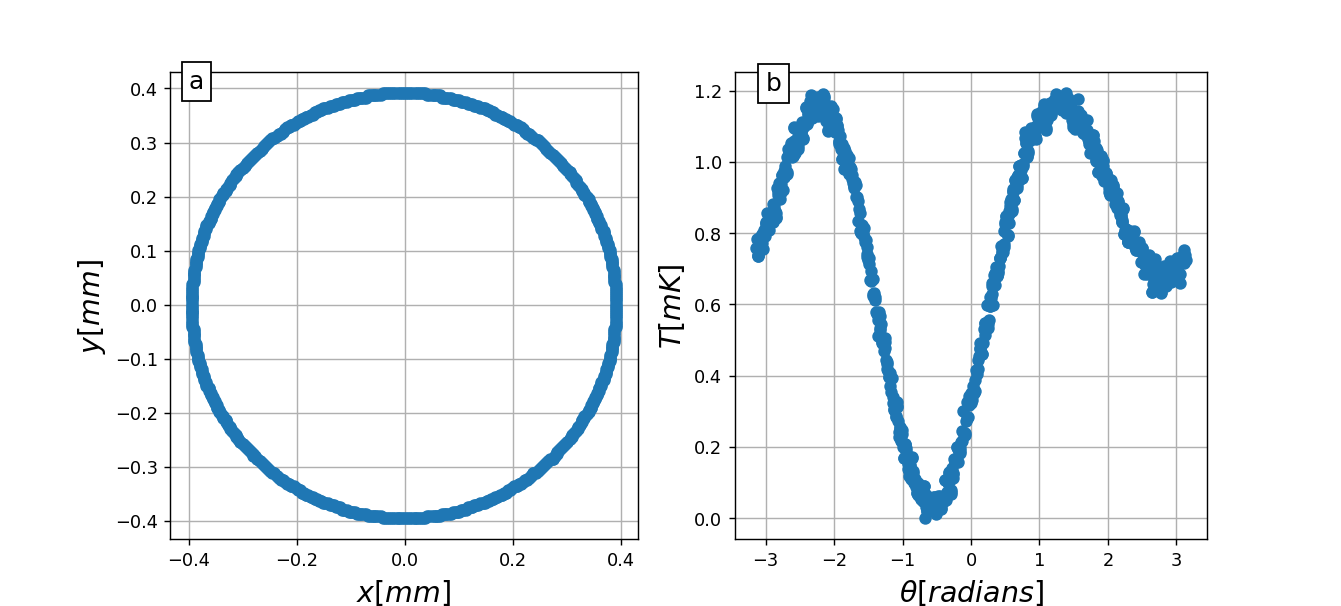}}
\caption{ a: Coordinates of the minima of the compensated MH potential. b: Pseudo-potential converted to a temperature along the minimum of the compensated MH potential. Trapping parameters indicated in text and  $\hat{r}_{min}$ tuned to 0.1. }
\label{fig:example_ring_temperature}
\end{figure}

In the compensated situation of Figure~\ref{fig:pseudo_example}, the exact positions of the calculated minima and their trapping potential have been evaluated along $\hat{r}_{min} = 0.1$ corresponding to $393\mu$m.
The results are presented in Figure~\ref{fig:example_ring_temperature}. In Figure~\ref{fig:example_ring_temperature}(a), the calculated positions of the potential minima are fitted by an ellipse. A value of 1.00010 is obtained for the degree of ellipticity and the ellipse can be considered as a perfect circle for most applications.
In Figure~\ref{fig:example_ring_temperature}(b), the variations of potential converted to temperature variations have been plotted. The corresponding temperature scale has been shifted to have $T_{min}=0$.
For this particular case, the observed variation along the minimum ring is $\Delta T = 1.19$~mK.

The homogeneity of the trapping potential can be tested with the variation of two critical parameters for the realisation  of the ion ring, which are the mechanical errors in the rod positioning and the desired working ring radius. The working ring radius plays a role in the expected final topology as a single ring structure of given size can only contain a limited number of ions. In order to increase the ion number in a single ring, the ring radius $\hat{r}_{min}$ has to be increased before a transition to two rings occurs.

\begin{figure}
\center
\scalebox{0.55}{\includegraphics{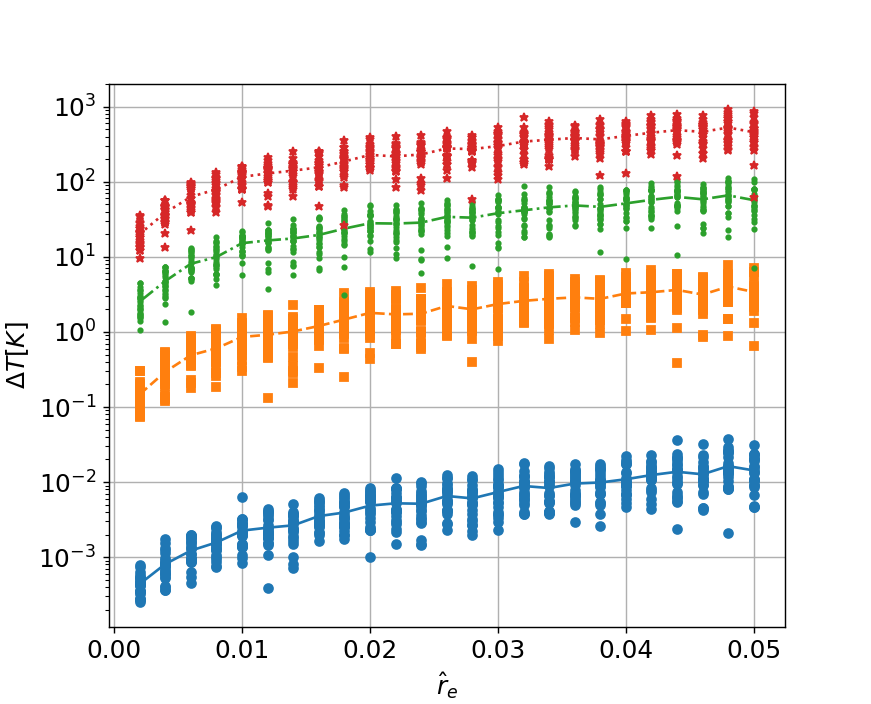}}
\caption{Potential inhomogeneity converted to temperature $\Delta T$, along the minimum MH ring reached through compensation, as a function of the mechanical error, for different values of $\hat{r}_{min}$. Blue: $\hat{r}_{min} =0.1$; orange: $\hat{r}_{min} = 0.2$; green: $\hat{r}_{min} = 0.3$; red: $\hat{r}_{min} = 0.4$. The marks at a given value of $\hat{r}_{e}$ correspond to the 25 random sets, while the lines represent the evolution of the mean values. }
\label{fig:temp_vps_error}
\end{figure}

The mechanical error is defined by the normalized amplitude $\hat{r}_{e} = r_{e} / r_{0}$ of the displacement of each rod in a random direction. For each value of $\hat{r}_{e}$ studied, 25 different random sets have been used, similarly to Pedregosa et al.~\cite{pedregosa18}. Figure~\ref{fig:temp_vps_error} shows the potential variation (converted to $\Delta T$) once the potential correction has been applied, as a function of the mechanical error, $\hat{r}_{e}$. Therefore, several values of $\hat{r}_{min}$ are also shown in Figure~\ref{fig:temp_vps_error}.

The expected ion temperature  in an octupole trap is of the order of 10~mK, resulting from the competition between RF heating and laser cooling \cite{marciante_structural_2012}. Figure \ref{fig:temp_vps_error} shows that in order to reach $\Delta T$ lower than this value, ion rings of $\hat{r}_{min} \le 0.1$ should be used in combination with low mechanical errors ($\left<\Delta T\right> = 5$~mK for $\hat{r}_{e}=0.02$ and $\hat{r}_{min} = 0.1$). In our set-up, $\hat{r}_{e}=0.02$ corresponds to a mechanical error of 80~$\mu$m.

\subsection{The role of the fitting area}
\begin{figure}
\center
\scalebox{0.55}{\includegraphics{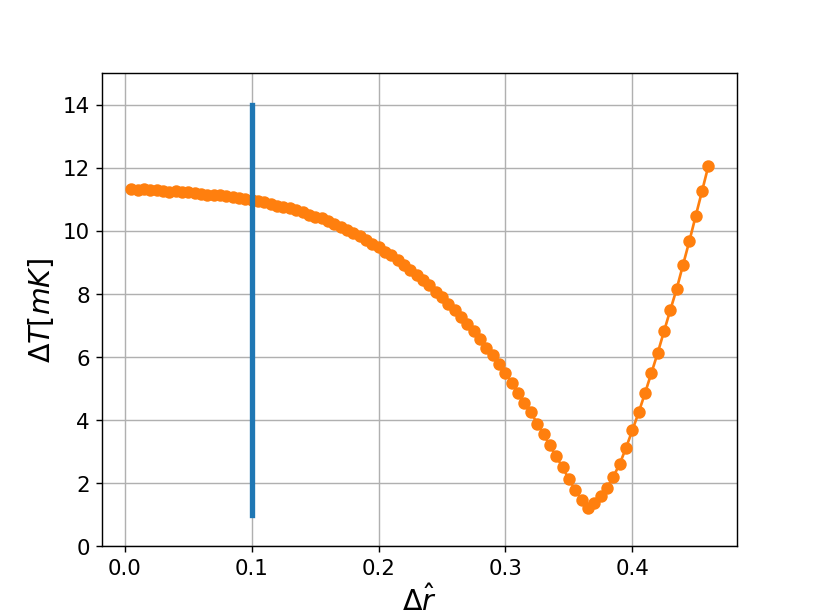}}
\caption{Typical evolution of the potential inhomogeneity converted to $\Delta T$  on the minimum MH ring versus the radial width of the surface used for the least-squares minimisation, defined by $\hat{r}_{ring} \pm \Delta r$. The vertical blue line corresponds to $\Delta r= r_{min}$ which is the limit between the ring and disk surface.}
\label{fig:example_ring_temperature_vs_width}
\end{figure}

As mentioned earlier,
the choice of the area used in the least-squares minimisation  plays an important role in the final results. Due to the symmetries of the problem, a ring of radius $r_{min}$ and half-width $\Delta r$ has been chosen as region to perform the double sum of equation~\ref{eq:chi2}. For $\Delta r>r_{min}$, this region becomes a disc with radius ($r_{min}+\Delta r$).
In the case of the misalignment configuration used for Figure~\ref{fig:pseudo_example} and for $\hat{r}_{min} = 0.1$, Figure~\ref{fig:example_ring_temperature_vs_width} shows the evolution of $\Delta T$ with $\Delta \hat{r}=\Delta r/r_{0}$. The vertical blue line corresponds to $\Delta r= r_{min}$.

The curve shows a strong dependency on the choice of the region considered. This graph seems to indicate that there is an optimum value of area for the compensation algorithm to obtain the lowest $\Delta T$ which corresponds to values which are at least 3 times larger than the ring radius of the MH potential. For smaller areas, it seems that there are not enough points that are taken into account to optimise the compensation on the MH ring; whereas for larger areas, it appears that the correction of defects deteriorates the optimisation along the MH ring. While understanding the exact reasons behind such behaviour would have been an interesting numerical problem, it would have been time consuming without bringing any new physics to the problem at hand an therefore it has not been undertaken.



For all curves presented in this article, the optimization of the compensation has been performed for the optimum area corresponding to the minimum value of $\Delta T$.

\subsection{The compensation voltage amplitudes}
Another important aspect is the amount of correction on $V^{RF}_{l}$ that is needed. This point is crucial to design the RF power supply.

\begin{figure}
\center
\scalebox{0.55}{\includegraphics{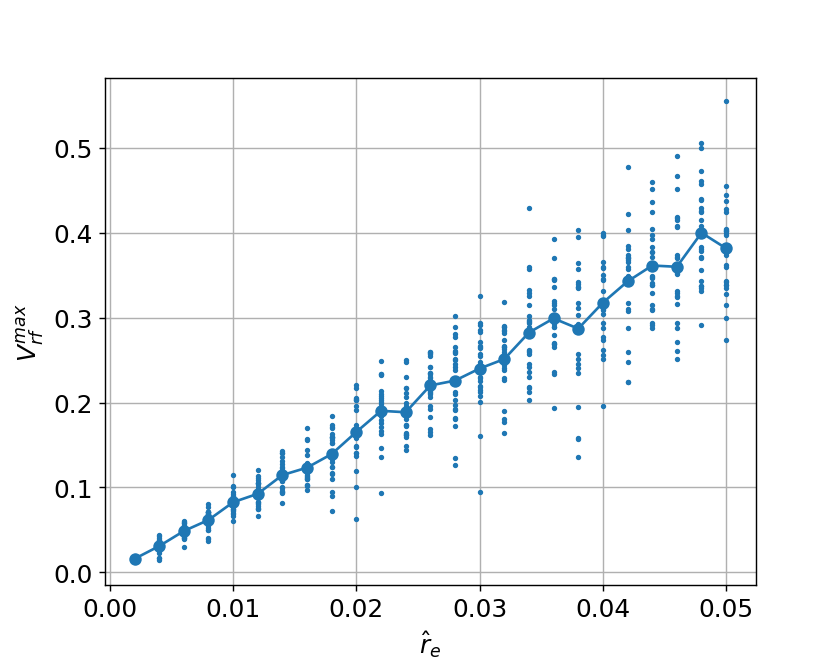}}
\caption{Relative difference between the minimum and maximum values for optimal compensation versus the relative mechanical positioning error, $\hat{r}_{e}$, same configuration as in figure~1.  The cloud of small marks at each value of $\hat{r}_{e}$ corresponds to the 25 different random sets of errors while the larger marks indicate the mean values. }
\label{fig:Vmax_vs_re}
\end{figure}

In the present implementation, the minimisation algorithm acts on eight electrodes. The difference between the highest and the lowest values relative to the nominal value, corresponding to the best compensation values found by the algorithm are plotted in figure~\ref{fig:Vmax_vs_re} as a function of the normalised mechanical error, $\hat{r}_{e}$.   In this example, the choice of $\hat{r}_{min}$ is not decisive as there is less than 2\% difference for different values of $\hat{r}_{min}$. We observe a linear relationship
which confirms the intuitive assumption that the larger the mechanical errors, the larger the required correction voltages.

\subsection{In terms of impact on the cold ions arrangement}

Molecular Dynamics (MD) simulations provide a virtual implementation of such a compensation for a particular experimental set-up, and allow to probe the influence of the corrected potential on the trapped ions. To this purpose, the code developed in~\cite{pedregosa15,pedregosa18}, has been modified to use the CPO generated matrices containing the electric fields, $E_{x}$, $E_{y}$ and a $5^{th}$ order interpolation is used to find the electric field at the ion position. The simulation parameters are: 128 Ca$^{+}$-ions, a nominal value of RF (before corrections) of $V_{RF} = 300$~V, $\Omega/2\pi = 3.5$~MHz, $\omega_{z}/2\pi = 100$ kHz.

The MD simulations uses an initial velocity re-scaling algorithm that cools down the ions to a temperature of 10~mK \cite{pedregosa15}. While such technique is not physically correct, allows to speed up the simulation and provides initial conditions for the last part of the simulation, when the code switches to a laser cooling algorithm that uses a two-level models for the laser-interaction through absorption and emission of photons \cite{marciante_structural_2012}

The final ion temperatures, calculated as the averaged temperature over the last 3000 RF periods ($\approx 0.9$~ms), are $T=9.6 \pm 0.7$~mK and $T=14.2 \pm 1.4$~mK for the compensated and uncompensated case respectively.

\begin{figure}
\center
\scalebox{0.55}{\includegraphics{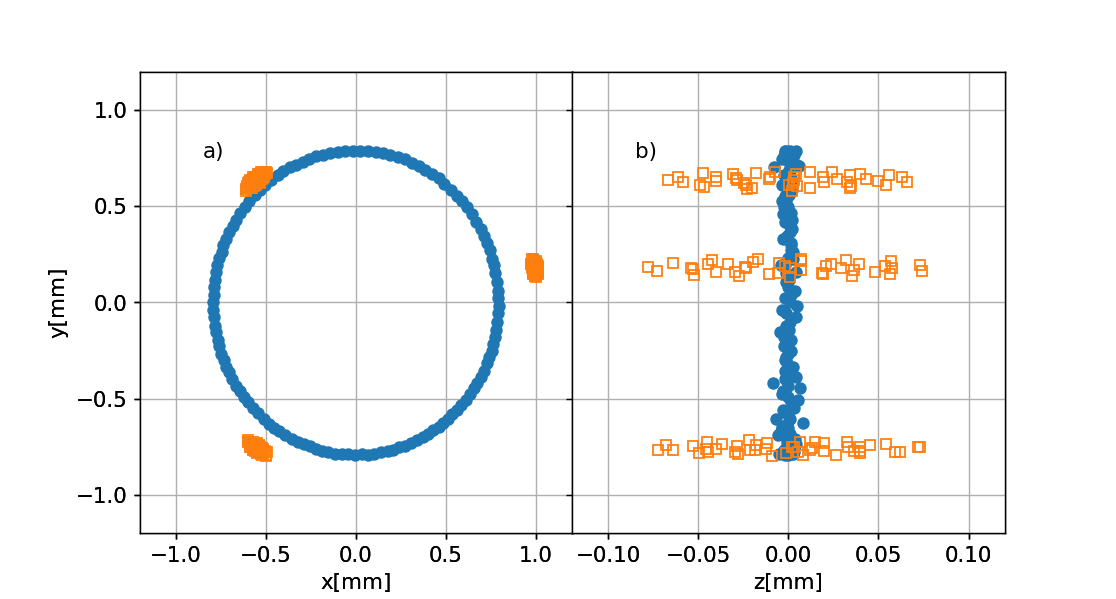}}
\caption{Ion positions for the two potentials shown in figure~\ref{fig:pseudo_example}. a) $x-y$ projection, b) $z-y$ projection. Full circles correspond to the compensated potential while the empty squares correspond to uncompensated potential. $r_{0}=3.93$~mm.}
\label{fig:XYZ_Before_After_Comp}
\end{figure}

Figure~\ref{fig:XYZ_Before_After_Comp}, shows the $x-y$ (left) and the $z-y$ (right) projection of the ion positions for two different cases, the compensated and uncompensated potentials corresponding to the two potentials shown in Figure~\ref{fig:pseudo_example}. We observe that the compensation permits to obtain an ion ring instead of three chains in the uncompensated three potential minima.

\section{Experimental implementation}

\subsection{Requirements for compensation voltages}
In the following we discuss the technical implementation of the proposed compensation method, and we detail the requirements concerning the resolution and stability of the RF trapping drive.

\subsubsection*{Resolution}
\begin{figure}
\center
\scalebox{0.6}{\includegraphics{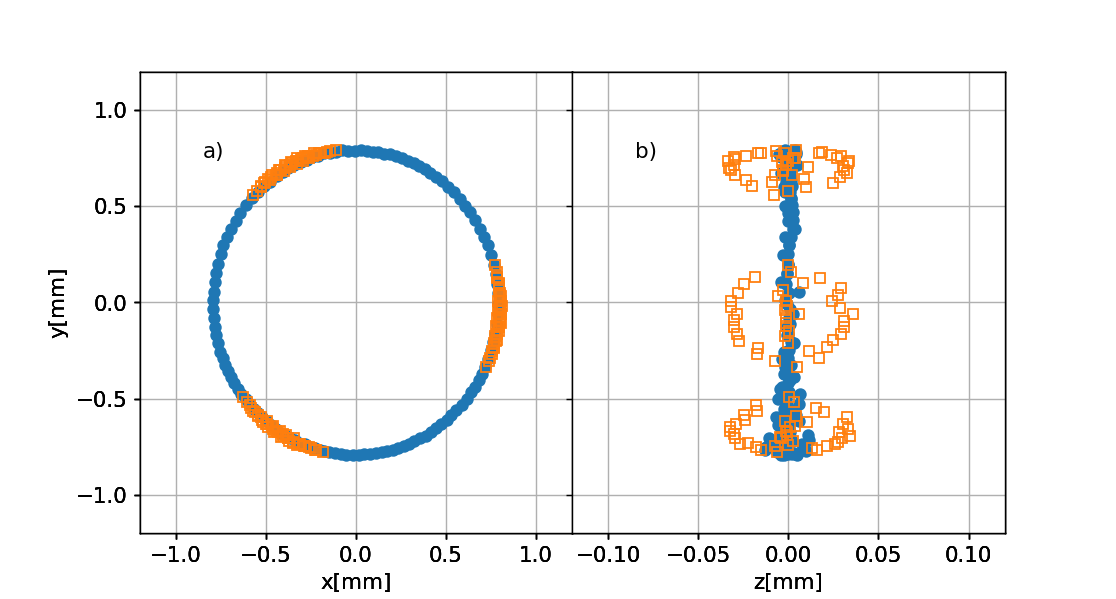}}
\caption{Final ion positions determined by MD simulation where a potential compensated with a noise-free power supply but limited resolution. a) x-y projection; b) z-y projection. RF voltages are defined with a resolution of $10^{-3}$ (blue full circles) and $10^{-2}$ (orange empty squares). Simulation parameters are indicated in the text.}
\label{fig:XY_Res1e2_vs_Res1e3}
\end{figure}

The results of Figure~\ref{fig:XYZ_Before_After_Comp} have been obtained with values of $V^{RF}_{l}$ defined with the usual numerical double precision (15 Significant Digits) when scaled to 1~V. Typical RF power supplies provide much lower resolution. In the following we present the effect of this reduced resolution on the final ion distribution by exploring the ion dynamics with MD simulations for the case studied in Figure~\ref{fig:pseudo_example}. A noise-free RF source is assumed. The different resolutions studied ranged from $10^{-2}$ to $10^{-5}$. As Figure~\ref{fig:XY_Res1e2_vs_Res1e3} shows, a resolution of $10^{-2}$ leads to three local minima while in the case of a $10^{-3}$ resolution, the ions form already a ring-like structure.

\begin{figure}
\center
\scalebox{0.6}{\includegraphics{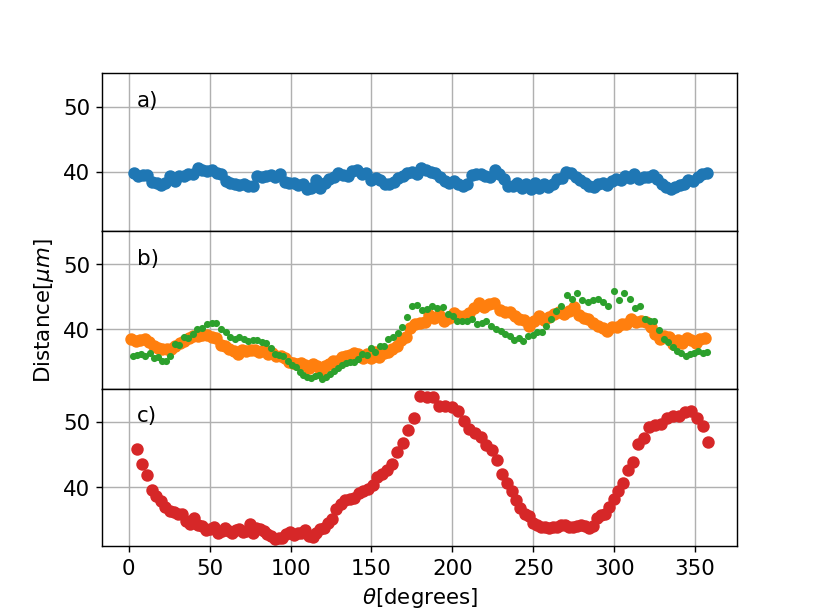}}
\caption{Ion-pair distance vs the ion-pair angular position for different resolutions. a) Error-free; b) 10$^{-15}$ (orange circles); 10$^{-4}$ (green dots); c) 10$^{-3}$. The other parameters of the MD simulation are like in Figure~\ref{fig:XYZ_Before_After_Comp}.}
\label{fig:InterIonDist_vs_Res}
\end{figure}

The homogeneity of the ion ring can be expressed as the distance between two consecutive ions as a function of the ion-pair angular position. In Figure~\ref{fig:InterIonDist_vs_Res}, such a curve is shown for several resolutions. The ion-pair distance has been obtained as follows: the average position over the last 100 RF periods is computed and used to obtain the distance (in 3D) to the left neighbour. Figure~\ref{fig:InterIonDist_vs_Res} shows that, for the case studied, there is only a very small difference between the full resolution (numerical double precision) and $10^{-4}$ (b),  both presenting a small deviation with respect to the error free case (a), while a significant difference is observed if the RF drive's resolution is lowered to $10^{-3}$ (c). The mean inter-ion distances and the final temperatures (average over the last 100 RF periods) are given in table \ref{tab:1}. The reduction of the resolution has no impact on the temperature of the ion sample but has a signature on the deviation of the inter-ion distance, which is nearly doubled when the RF voltages  are  defined with a resolution lowered from $10^{-4}$ to $10^{-3}$. As this increase cannot be imputed to a thermal effect, we conclude it is induced by the profile of the pseudo-potential. Inter-ion mean distance and deviation should be compared to the average separation of 9 $\mu$m reported in  Tabakov {\it et al.}~\cite{tabakov15}, for a  ring structure of approximately 400 ions, trapped in a micro-fabricated quadrupole circus.  The  corresponding inter-ion distance deviation can be estimated to 10\% of its mean value, as deduced from Figure 5~\cite{tabakov15}.
\begin{table}
\begin{center}
\begin{tabular}{l|c|c|c|c}
\hline{}
resolution & Error-Free & $10^{-15}$& $10^{-4}$ & $10^{-3}$\\
\hline
Inter-Ion Distance & $38.9 \pm 0.9 \mu$m & $38.9 \pm 2.7 \mu$m & $38.9 \pm 3.5 \mu$m & $39.2 \pm 6.7 \mu$m \\
\hline{}
$<T>$ & $5.7\pm0.8 $ mK & $5.5\pm0.8 $ mK & $5.5\pm0.8$ mK & $5.5\pm0.7 $ mK \\
\hline
\end{tabular}
\end{center}
\caption{Results from MD simulations of 128 laser-cooled ions organised in a ring, in an error-free octupole trap, and in a real octupole trap as defined in the text. The computed default compensating RF voltages are defined with different resolution for each simulation.}\label{tab:1}
\end{table}

A minimum resolution of the order of $10^{-4}$ seems therefore necessary for compensation in the studied case. This represents a voltage resolution of $\approx 30$~mV for a typical RF voltage of 300~V. This imposes a strict, but feasible constraint on the RF source.

\subsubsection*{Stability}
The limited resolution of RF supplies translates into a static mismatch between the desired and the applied RF voltages and has a signature on the ion organisation in the ring. In a second step, as the trapping is a dynamical process, we expect an eventual  noise on the applied RF voltages to have an impact on the ion temperature. To quantify this impact,  noise on the RF voltage amplitude is introduced in the MD code around the nominal compensated values: $V^{noise}_{l} = V^{RF}_{l}(1 + \beta\cdot rand)$, where $\beta$ is the amplitude of the noise and $-1 < rand < +1$. The values applied for $V^{RF}_{l}$ have a $10^{-4}$ resolution. It is implemented in two different ways:  random values are taken every time step ($dt = 2\pi /(100 \Omega)$)or, at every time step, a random number decides if an update on $V^{noise}_{l}$ needs to be taken or not with equal probability. If decided, the update on $V^{noise}_{l}$ is then randomly taken.
The simulations show that the final temperature of the ions increases by less than a factor of 2 for noise going from $\beta = 10^{-4}$ to $10^{-1}$.
From this result it is clear that the noise implemented does not seem to have a significant impact on the ion velocity distribution. The reason for this immunity to noise could be explained by the fact that  only the large noise contribution at peak amplitude significantly affects  the dynamics. This occurs only during a very small fraction of the RF period. Therefore, a white noise of amplitude lower than $10^{-1}$ 
should be a sufficient condition for compensation, which is well in the range of usual specifications for RF sources.


\subsection{Practical implementation of diagnosis}
In a real experimental scenario, diagnostic techniques are needed to find the right values of $V^{RF}_{l}$ as the actual error for each rod will not be known. An experimentally feasible approach is to start with equal voltage on all of the electrodes, so that the ions organise themselves in three clouds/crystals at the position of the three local minima, as shown numerically in figure~\ref{fig:XYZ_Before_After_Comp} and experimentally in figure 1 of Pedregosa et al.~\cite{pedregosa18} Assuming that the distances between  the centre of the three  ion clouds can be deduced from fluorescence imaging, their sum could be used as an input in a non-linear algorithm which remotely control the RF amplitude of each rod, searching for the minimisation of this quantity.

In order to test such an approach, a Python code has been developed using the non-linear minimization algorithm (Nelder-Mead type) from the Scipy library~\cite{gao12}. It computes the pseudo-potential depending on the free parameters ($V^{RF}_{l}$) using eq.~\ref{eq:experimental_octo}, finds the position of all the relative minima and computes the distance between them. The sum of the three distances  is the quantity minimised by the algorithm. When applied to the case corresponding to our experimental set-up (see figure~\ref{fig:pseudo_example} ) the code converges after 576 function evaluations. The compensated pseudo-potential along the ring shows deviations of $\Delta T = 3.6$~mK (for $\hat{r}_{min}=0.1$) when converted into temperature. In a second step, the standard deviation of the inter-ion distance should serve as the input parameter to be minimized to reach an homogeneously distributed ion ring.

\section{Conclusion}
The present work shows that deviations from the ideal positioning of the rods forming an octupole RF trap can be corrected by using individually adapted RF amplitudes to each trap rod. In order to do so, a least-squares minimisation approach has been implemented for the multipole trapping potential. The evaluation of the goodness of this solution gives an overview of the performances that can be reached, and imposes technical conditions which are strict but technically feasible. The choice of the RF power supply requires a minimum resolution in amplitude of at least $10^{-4}$.

The present results demonstrate the feasibility of realising an ideal multipole potential, which could be a solution to overcome the incomplete rotational cooling which has been reported before \cite{endres17}. It is also  the necessary condition to generate homogeneous ion rings which are of interest for different high precision applications. 
Various experimental approaches are pursued today in order to be able to create these structures with outstanding symmetry performances, which allow to imagine novel measurements with characteristics that the best individual ion chains cannot reach. Moreover, only linear multipole RF traps will eventually allow the generation of ion crystal tubes, representing an interesting example of 2D structures with periodic boundary conditions.

\section*{Acknowledgements}
JPG thanks S. Schlemmer for fruitful discussions. This work has been financially supported by ANR (ANR-08-JCJC-0053-01),
CNES (contracts no. 116279 and 151084), and R\'egion PACA. MRK
acknowledges financial support from CNES and R\'egion
Provence-Alpes-C\^ote d'Azur.





\end{document}